# A low field technique for measuring magnetic and magneto-resistance anisotropy coefficients applied to (Ga,Mn)As.

J A Haigh, A W Rushforth, C S King, K W Edmonds, R P Campion, C T Foxon, and B L Gallagher

School of Physics and Astronomy, University of Nottingham, University Park, Nottingham, NG7 2RD United Kingdom

*We demonstrate a simple, low cost, magneto-transport method for rapidly characterizing the magnetic anisotropy and anisotropic magneto-resistance (AMR) of ferromagnetic devices with uniaxial magnetic anisotropy. This transport technique is the analogue of magnetic susceptibility measurements of bulk material but is applicable to very small samples with low total moment. The technique is used to characterize devices fabricated from the dilute magnetic semiconductor (Ga,Mn)As. The technique allows us to probe the behavior of the parameters close to the Curie temperature, in the limit of the applied magnetic field tending to zero. This avoids the complications arising from the presence of paramagnetism.*

The characterization of the magnetic properties of bulk ferromagnetic materials has traditionally involved direct measurements of the magnetization using instruments such as superconducting quantum interference device (SQUID) magnetometers or vibrating sample magnetometers (VSM). Electrical transport measurements can yield the magneto-transport coefficients and have also been used to study the magnetic anisotropies of conducting ferromagnetic metals[1] and semiconductors[2,3] in individual micron sized samples, where direct magnetometry techniques would not be sufficiently sensitive. The dilute magnetic semiconductor (Ga,Mn)As has revealed a rich behaviour of the magnetotransport coefficients[4,5] and the magnetic anisotropy[6,7] owing to the strong spin orbit coupling of the charge carriers in this system. Recently, a detailed transport investigation was used to extract the magnetic anisotropy coefficients of (Ga,Mn)As samples by performing high resolution field sweeps over a range of angles and to fields in excess of the anisotropy fields[3]. This technique provides an accurate determination of the magnetic anisotropy, but requires extensive and detailed measurements to achieve the desired resolution. Here, we present a relatively simple transport technique which enables the magnetic anisotropy and the magnetotransport coefficients to be extracted to a good accuracy. In addition, it allows the magnetotransport coefficients to be extracted at temperatures close to the Curie temperature ($T_C$) in the limit where the applied magnetic field tends to zero, thereby avoiding the complicating effects of induced paramagnetic moments[8].

Our technique involves the application of very small alternating square wave in-plane magnetic field pulses to a sample with uniaxial magnetic anisotropy which shows single domain behavior, and measuring the induced changes in the longitudinal and transverse anisotropic magnetoresistance (AMR). The amplitude of the field pulses is much smaller than the anisotropy field and produces small perturbations of the direction of the magnetization. The technique is similar to the "AC-AMR" technique introduced by Venus et al[9] which is the transport analogue of ac-magnetic susceptibility measurements.

However in our case we measure the response essentially at equilibrium. This simplifies interpretation but yields no information on dissipation.

The measurements were carried out at temperatures down to 1.5K on a Hall bar sample with dimensions 235μm x 45μm, fabricated from a 25nm $Ga_{0.94}Mn_{0.06}As$ epilayer grown by low temperature molecular beam epitaxy on a GaAs(001) substrate and buffer layers. SQUID magnetometry measurements show that this material has a Curie temperature of 84K and that the magnetization is in-plane at all temperatures with a dominant uniaxial anisotropy along the $[1\bar{1}0]$ direction at temperatures above 20K. The sample was fabricated so that the current direction was close to the $[1\bar{1}0]$ easy axis direction. In practice we found a misalignment angle $\theta_0$ of order $2°$. The magnetic field was provided by an air cored solenoid within μ-metal shielding. Background/offset magnetic fields were ≈0.05mT, small enough to be neglected in our analysis. The sample was rotated with the applied field in the plane of the Hall bar.

The AMR of this (Ga,Mn)As sample is well approximated[10] by $\Delta R_{xx}=\Delta R\cos2\theta$ and $\Delta R_{xy}= \Delta R\sin2\theta$ where $\Delta R_{xx}$ and $\Delta R_{xy}$ are respectively the changes in the longitudinal and transverse resistances when the magnetization (M) makes an angle θ to the direction of the current. $\Delta R = R_{\parallel} - R_{\perp}$ is the difference in $R_{xx}$ for M parallel ($R_{\parallel}$) and perpendicular ($R_{\perp}$) to the current direction. The application of positive ($H^+$) and negative ($H^-$) in plane field pulses perturbs M by a small angle in the film plane. Figures 1(a) and (b) show a typical sequence of applied field pulses and the corresponding response of $R_{xy}$. The dwell time at each applied field is much longer that the time constant of the measurement system which is ~ 2 ms. Figure 1(c) shows a typical response of the measured $R_{xy}$ as the field pulses are applied at a series of in plane directions. The response shows clear uniaxial behaviour with zero response for fields applied along the easy axis (which we take as $\varphi = 0$) and maximum response for $\varphi$ close to $\pm 90^0$. Thus, these simple measurements give a direct measurement of the easy axis direction[11].

For the applied field fixed along the hard axis ($\varphi = \pm 90^0$) small positive ($H^+$) and negative ($H^-$) field pulses produce small deviations $\pm\Delta\theta$ of the magnetization from the easy axis direction and the resulting changes in resistance are

$$\Delta R_{xx}^{\pm} = \Delta R[\cos 2(\theta_0 \pm \Delta\theta) - \cos 2\theta_0] \quad \text{and} \quad \Delta R_{xy}^{\pm} = \Delta R[\sin 2(\theta_0 \pm \Delta\theta) - \sin 2\theta_0] \qquad (1)$$

$\theta_0$ and $\Delta\theta$ can be determined from $\tan\Delta\theta = -\Sigma\Delta R_{xx}/\Delta\Delta R_{xy}$ and $\tan 2\theta_0 = -\Delta\Delta R_{xx}/\Delta\Delta R_{xy}$, where $\Sigma\Delta R_{xx/xy} = \Delta R_{xx/xy}^{+} + \Delta R_{xx/xy}^{-}$ and $\Delta\Delta R_{xx/xy} = \Delta R_{xx/xy}^{+} - \Delta R_{xx/xy}^{-}$. The AMR coefficient $\Delta R$ can then be obtained from (1). Writing the magnetic free energy as

$$E/M = H_U \sin^2 \Delta\theta - H_C/4 \sin^2 2\Delta\theta - H\cos(\varphi - \Delta\theta) \qquad (2)$$

where $H_U$ and $H_C$ are the uniaxial and cubic magnetic anisotropy fields, and minimizing with respect to $\Delta\theta$ gives $H_U - H_C = H\sin(\varphi - \Delta\theta)/\sin 2\Delta\theta$ for $\Delta\theta^2 \ll 1$.

Figures 2(a), (b) and (c) show the measured temperature dependence of $\Delta R_{xx}^{\pm}$, $\Delta R_{xy}^{\pm}$ and $\Delta\theta$ for field pulse amplitude 3.7mT. The fact that $|\Delta R_{xx}^{+}| \neq |\Delta R_{xx}^{-}|$ reflects the non-zero value of $\theta_0$, the small misalignment of the current direction with respect to the $[1\bar{1}0]$. We extract $\theta_0 \approx 2°$ from our analysis. $\Delta\theta$ increases as the temperature approaches $T_C$, where the anisotropy energy becomes small relative to the applied field. Δθ also increases and has a maximum below ≈20K. This arises from a spin reorientation transition (SRT)[12] at the crossover from uniaxial [1$\bar{1}$0] to cubic [100]/[010] anisotropy. The difference in anisotropy fields, $H_U$-$H_C$ extracted using the method described above, is shown in Fig. 2(d). The shaded band indicates the level of uncertainty which is dominated by neglect of the crystalline AMR coefficients[10], which lead to small differences between the longitudinal and transverse values. Also shown are the $H_U$ and $H_C$ fields obtained by fitting the hard axis hysteresis loops[12] of the unprocessed (Ga,Mn)As film, measured by SQUID magnetometry. The good agreement between values of $H_U$-$H_C$ obtained from the two methods demonstrates the validity of our simple technique.

Figure 3 shows the low field AMR coefficient obtained from the data of figure 2 compared to values obtained from fitting to plots of resistance versus angle using a large magnetic field of 1T to overcome the magnetic anisotropy. The two types of measurement are in reasonable agreement at low temperatures. However, the low field AMR coefficient falls to zero close to $T_C$ while the values obtained from the high field data are finite well above $T_C$. A recent study[8] observed strong field and temperature dependences of the cos2θ component of the AMR in a (Ga,Mn)As sample for temperatures above and below $T_C$. A two-phase model, involving separate ferromagnetic and super-paramagnetic phases below $T_C$, was invoked to explain this behavior. The method presented here allows us to study the behavior of the AMR in the limit where the external magnetic field tends to zero, and so obtain the AMR associated with only the ferromagnetic phase. In this limit the temperature dependence of the AMR coefficient is similar to that of the spontaneous magnetization, falling toward zero as the temperature approaches $T_C$. The AMR above $T_C$ resulting from the application of a field of 1T has a temperature dependence similar to the measured induced magnetization. Below $T_C$, the AMR and magnetization measured at 1T and at zero field show similar temperature dependence. This indicates that, for our samples the observed field-dependence of the AMR is consistent with normal single phase behavior and is associated with the large susceptibility of ferromagnetic materials close to $T_C$.

In summary, we have demonstrated a simple, low cost, magneto-transport technique to characterize the magnetic anisotropy and AMR transport coefficient of a micron sized ferromagnet device with uniaxial magnetic anisotropy. The application of the technique to (Ga,Mn)As allows us to study the behavior of these parameters in the limit as the field tends to zero, and to separately identify the effects of ferromagnetism and paramagnetism on the AMR coefficient.

We acknowledge funding from EU NAMASTE grant 214499 and EPSRC grant GR/S81407/01.

**Figure 1 (Color online)** (a) A typical magnetic field pulse profile and (b) the measured resulting transverse resistance. (c) Angular dependence of $\Delta R_{xy}$ for applied field (1.8 mT) in the plane of the sample.

**Figure 2 (Color online)** The temperature dependence of (a) $\Delta R_{xx}$, (b) $\Delta R_{xy}$, (c) $\Delta\theta$ and (d) the anisotropy field $H_U$-$H_C$ measured using hard axis field pulses of ±3.7mT. The anisotropy fields extracted from fitting to the hard axis SQUID magnetometry data are also shown in (d). The shaded area allows for a 10% variation in the coefficient $\Delta R$ between $\Delta R_{xx}$ and $\Delta R_{xy}$ in equation (1)[10].

**Figure 3 (Color online)** Left axis: The temperature dependence of the AMR coefficient extracted from the field pulse experiment. The second order coefficient extracted from fitting to the data obtained in a field of 1T is also shown for comparison. The AMR coefficient is defined as $\Delta R / R_{av}$, where $R_{av}$ is the average resistance when the magnetization is rotated through 360° in the plane of the device. Note that for the low field data $R_{av}$ cannot be measured directly. Instead we use $R_{av}=R(B=0T)+\Delta R$, since the magnetic easy axis is parallel to the current at B=0T. Right axis: The magnetization measured by SQUID magnetometry along the $[1\bar{1}0]$ easy axis direction in a field of 1T (red circles) and the remnant magnetization, $M = \{M^2_{[1\bar{1}0]} + M^2_{[110]}\}^{\frac{1}{2}}$, where $M_{[1\bar{1}0]}$ and $M_{[110]}$ are the magnetization measured along the $[1\bar{1}0]$ or $[110]$ directions respectively (black squares).

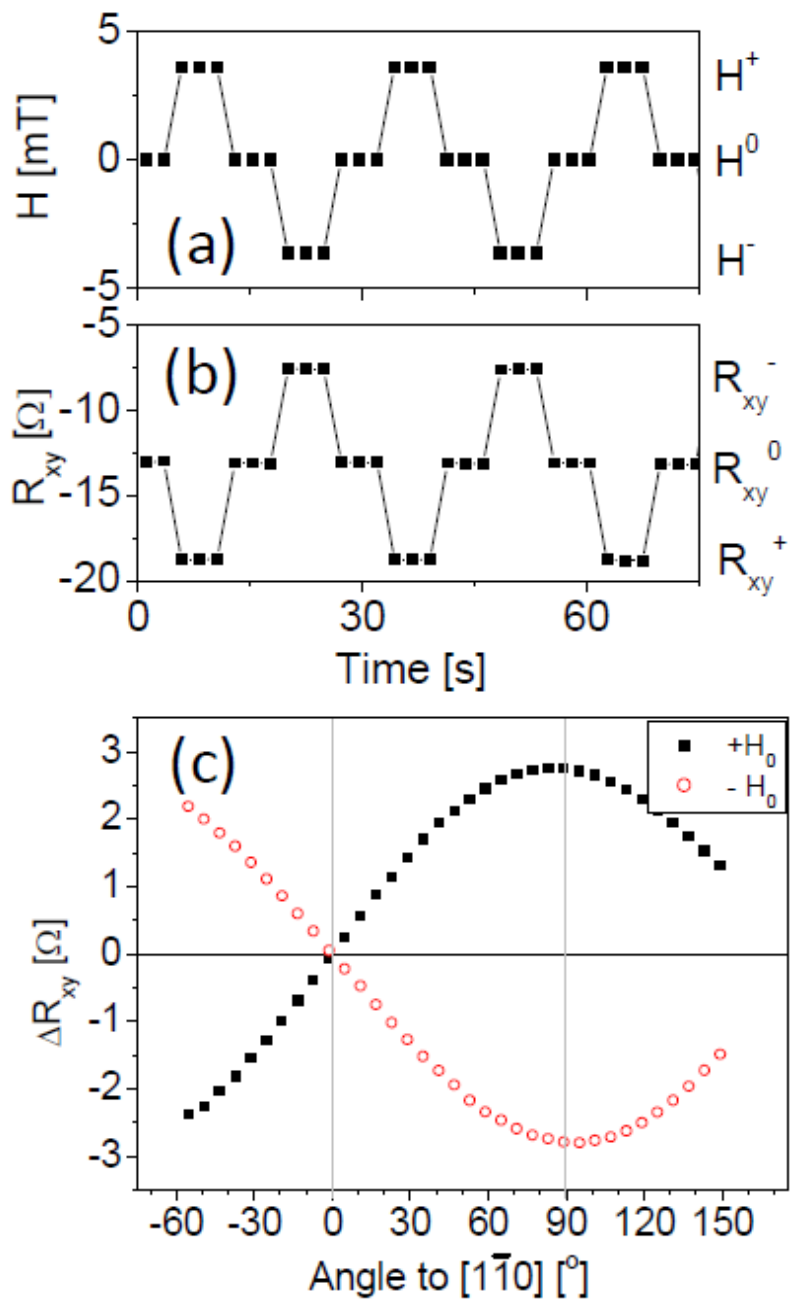

Figure 1

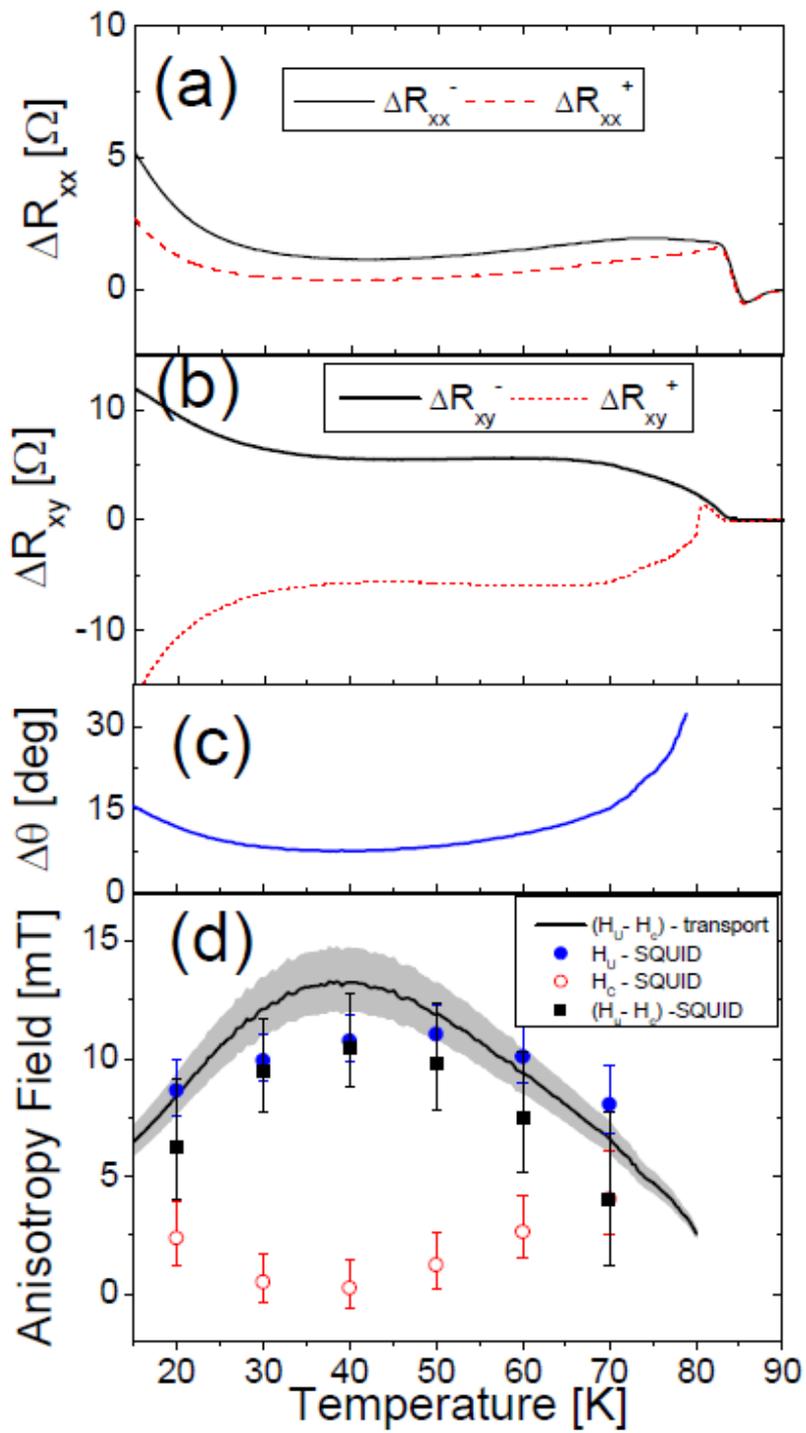

Figure 2

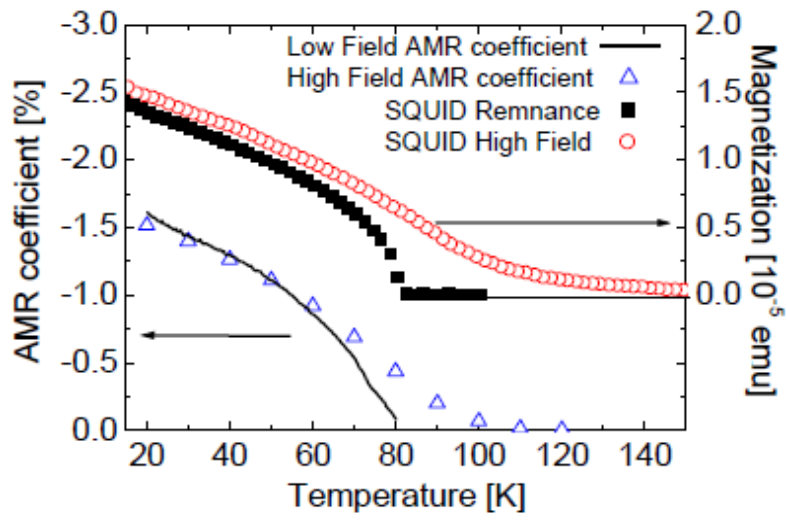

Figure 3

[1] E Dan Dahlberg, Kevin Riggs, and G A Prinz, J. Appl. Phys. **63**, 4270 (1988); K T Riggs, E Dan Dahlberg, and G A Prinz, Phys. Rev. B **41**, 7088 (1990).

[2] K Y Wang, K W Edmonds, R P Campion, L X Zhao, C T Foxon, and B L Gallagher, Phys. Rev. B. **72**, 085201 (2005).

[3] K Pappert, S Humpfner, J Wenisch, K Brunner, C Gould, G Schmidt, and L M Molenkamp, Appl. Phys. Lett. **90**, 062109 (2007); K Pappert, C Gould, M Sawicki, J Wenisch, K Brunner, G Schmidt, L W Molenkamp, New J. Phys. 9, 354 (2007).

[4] A W Rushforth, K Vyborny, C S King, K W Edmonds, R P Campion, C T Foxon, J Wunderlich, A C Irvine, P Vasek, V Novak, K Olejnik, Jairo Sinova, T Jungwirth, and B L Gallagher, Phys. Rev. Lett. **99**, 147207 (2007).

[5] H X Tang, R K Kawakami, D D Awschalom, and M L Roukes, Phys. Rev. Lett. **90**, 107201 (2003).

[6] M Sawicki, K Y Wang, K W Edmonds, R P Campion, C R Staddon, N R. S Farley, C T Foxon, E Papis, E Kaminska, A Piotrowska, T Dietl, and B L Gallagher, Phys. Rev. B **71**, 121302(R) (2005).

[7] T Dietl, H Ohno, and F Matsukura, Phys. Rev. B **63**, 195205 (2001); M Abolfath, T Jungwirth, J Brum, and A H MacDonald, Phys. Rev. B **63**, 054418 (2001).

[8] D Wu, Peng Wei, E Johnstone-Halperin, D D Awschalom, and Jing Shi, Phys. Rev. B. **77**, 125320 (2008).

[9] D Venus, F Hunte, I N Krivorotov, T Gredig, and E Dan Dahlberg, J. Appl. Phys. **93**, 8609 (2003); D Venus, F Hunte, and E Dan Dahlberg, J. Mag. Mag. Mat. **286**, 191 (2005).

[10] This is true for polycrystalline samples. For crystalline samples the AMR contains higher order components and depends also on the angle of the magnetization with respect to the crystallographic directions. A detailed study of the AMR for this sample[4] confirms that the approximation used here is valid to within 10%.

[11] From these measurements we can find the easy axis, $\theta = 0$, within $\sim 1^0$ and demonstrate that this axis is within $\sim 3^0$ of the $[1\bar{1}0]$ axis.

[12] K Y Wang, M Sawicki, K W Edmonds, R P Campion, S Maat, C T Foxon, B L Gallagher, and T Dietl, Phys. Rev. Lett. **95**, 217204 (2005).